\documentclass{aa}  
\usepackage{graphicx}
\usepackage{txfonts}
\usepackage{changes}
\usepackage{xcolor}
\begin{document}

   \title{First-year ion-acoustic wave observations in the solar wind by the RPW/TDS instrument on board Solar Orbiter}
   \titlerunning{Ion-acoustic waves observed by SolO RPW/TDS}
   \subtitle{}
   \author{D. P\'{i}\v{s}a
          \inst{1},
           J. Sou\v{c}ek\inst{1}, O. Santol\'{i}k\inst{1,2}, M. Hanzelka \inst{1,2}, G. Nicolaou \inst{3,4} \and
                        M. Maksimovic\inst{5} \and S.D. Bale\inst{6,7}, T. Chust\inst{8} \and Y. Khotyaintsev\inst{9} \and
                        V. Krasnoselskikh\inst{10} \and M. Kretzschmar\inst{10,11}, E. Lorf\`evre\inst{12} \and D. Plettemeier\inst{13}, 
                        M. Steller\inst{14}, \v{S}. \v{S}tver\'ak\inst{15,1}, P. Tr\'avn\'i\v{c}ek\inst{6,15}, 
                        A. Vaivads\inst{9,16}, A. Vecchio\inst{5,17}, T. Horbury \inst{18}, H. O'Brien \inst{18}, V. Evans \inst{18}, V. Angelini \inst{18}, C. J. Owen \inst{4} \and P. Louarn \inst{19}
                        }
          
   \authorrunning{D. Píša et al.}     
   \institute{Institute of Atmospheric Physics of Czech Academy of Sciences,
              Bocni II 1401, 141 00 Prague, \email{dp@ufa.cas.cz}
    \and Faculty of Mathematics and Physics, Charles University, V Holesovickach 2, 18000 Prague 8, Czech Republic
        \and Southwest Research Institute, San Antonio, TX 78238, USA
        \and Mullard Space Science Laboratory, University College London, Holmbury St. Mary, Dorking, Surrey RH5 6NT, UK
        \and LESIA, Observatoire de Paris, Universit\'e PSL, CNRS, Sorbonne Universit\'e, Univ. Paris Diderot, Sorbonne Paris Cit\'e, 5 place Jules Janssen, 92195 Meudon, France 
        \and Space Sciences Laboratory, University of California, Berkeley, CA, USA
        \and Physics Department, University of California, Berkeley, CA, USA
        \and LPP, CNRS, Ecole Polytechnique, Sorbonne Universit\'{e}, Observatoire de Paris, Universit\'{e} Paris-Saclay, Palaiseau, Paris, France
        \and Swedish Institute of Space Physics (IRF), Uppsala, Sweden  
        \and LPC2E, CNRS, 3A avenue de la Recherche Scientifique, Orl\'eans, France 
        \and Universit\'e d'Orl\'eans, Orl\'eans, France 
        \and CNES, 18 Avenue Edouard Belin, 31400 Toulouse, France
        \and Technische Universität Dresden, Würzburger Str. 35, D-01187 Dresden, Germany 
        \and Space Research Institute, Austrian Academy of Sciences, Graz, Austria   
        \and Astronomical Institute of the Czech Academy of Sciences, Prague, Czechia 
        \and Department of Space and Plasma Physics, School of Electrical Engineering and Computer Science, Royal Institute of Technology, Stockholm, Sweden
        \and Radboud Radio Lab, Department of Astrophysics, Radboud University, Nijmegen, The Netherlands
        \and Department of Physics, Imperial College, SW7 2AZ London, UK
        \and Institut de Recherche en Astrophysique et Planétologie, 9, Avenue du Colonel ROCHE, BP 4346, 31028 Toulouse Cedex 4, France
             }
   \date{}

   \abstract
{Electric field measurements of the Time Domain Sampler (TDS) receiver, part of the Radio and Plasma Waves (RPW) instrument on board Solar Orbiter, often exhibit very intense broadband wave emissions at frequencies below 20~kHz in the spacecraft frame. During the first year of the mission, the RPW/TDS instrument was operating from the first perihelion in mid-June 2020 and through the first flyby of Venus in late December 2020.}
   {In this paper, we present a year-long study of electrostatic fluctuations observed in the solar wind at an interval of heliocentric distances from 0.5 to 1~AU. The RPW/TDS observations provide a nearly continuous data set for a statistical study of intense waves below the local plasma frequency.}
   {The on-board and continuously collected and processed properties of waveform snapshots allow for the mapping plasma waves at frequencies between 200~Hz and 20~kHz. We used the
triggered waveform snapshots and a Doppler-shifted solution of the dispersion
relation for wave mode identification in order to carry out a detailed spectral and polarization analysis. }
   {Electrostatic ion-acoustic waves are the most common wave emissions observed between the local electron and proton plasma frequency by the TDS receiver during the first year of the mission. The occurrence rate of ion-acoustic waves peaks around perihelion at distances of 0.5~AU and decreases with increasing distances, with only a few waves detected per day at 0.9~AU. Waves are more likely to be observed when the local proton moments and magnetic field are highly variable. A more detailed analysis of more than 10000 triggered waveform snapshots shows the mean wave frequency at about 3~kHz and wave amplitude about 2.5~mV/m. The wave amplitude varies as $\mathrm{R^{-1.38}}$ with the heliocentric distance. The relative phase distribution between two components of the E-field projected in the Y-Z Spacecraft Reference Frame (SRF) plane shows a mostly linear wave polarization. Electric field fluctuations are closely aligned with the directions of the ambient field lines. Only a small number (3\%) of ion-acoustic waves are observed at larger magnetic discontinuities.}
   {}

   \maketitle
%
\section{Introduction}
The solar wind, defined as a super-sonic flow of plasma, originates at the Sun's corona and fills the entire Solar System. Processes in this dynamic environment give rise to plasma waves that interact with particles and modify their velocity distributions. The importance of plasma waves in the thermal stabilization of solar wind plasmas has been widely accepted \citep[e.g.,][]{marsch91}. At frequencies below the plasma frequency ($f_{pe}$), there are only two electrostatic modes with a wave vector that is parallel to the ambient magnetic field line, namely: ion-acoustic and electron-beam modes \citep{gurnett91}.

Early observations of the Helios I \& II spacecraft at heliocentric distances between 0.3~and 1~AU showed broadband electrostatic waves at frequencies below the local electron plasma frequency \citep{gurnett77,gurnett78}. The typical wave amplitude was 1~mV/m at 0.3~AU and decreased as $\sim$1/R with increasing heliocentric distances. The electric field strength positively correlated with the electron to ion temperature ratio, $T_e/T_i$ , and the electron heat flux \citep{gurnett91}. The high-resolution spectral measurements by the Voyager spacecraft showed that broadband electrostatic fluctuations measured by Helios were electrostatic fluctuations, with peak frequency changes on timescales on the order of a second \citep{kurth79}. 

These fluctuations were identified as ion-acoustic oscillations, which are strongly Doppler-shifted into the frequency range of $f_{pi} < f < f_{pe}$, where $f_{pi}$ and $f_{pe}$ are proton and electron plasma frequency, respectively. Possible generation mechanisms were suggested as the ion-beam and electrostatic electron-ion (electron heat flux) instabilities \citep{lemons79,gary87,gary91}.

Previous wave observations from missions such as Helios 1 \& 2 were limited to spectral measurements at heliocentric distances larger than 0.3~AU. The recent observations of the Parker Solar Probe mission \citep{fox16} provide both spectral and electric field waveforms in the region with a heliocentric distance of $\sim$0.26~AU from the Sun. \cite{mozer20b} presented observations of very intense broadband fluctuations in the frequency range from 100~Hz to tens of kHz in the spacecraft frame. Their detailed analysis showed that these fluctuations were electrostatic and linearly polarized. Waves were identified as the electrostatic ion-acoustic mode and are observed during several second-long bursts with amplitudes of 15~mV/m. Wave vectors are oriented anti-parallel to the ambient magnetic field. Based on ion measurements, they suggested that ion-beam instabilities more likely than current-driven instabilities to produce observed ion-acoustic waves.

This paper presents the first-year observations of intense ion-acoustic waves at frequencies between 200~Hz and 20~kHz by the Time Domain Sampler (TDS) receiver, part of the Radio and Plasma Waves (RPW) instrument on board Solar Orbiter. The observed waves are identified as the electrostatic ion-acoustic mode exhibiting linear polarization and wave vector parallel or anti-parallel with the local magnetic field. They are strongly Doppler-shifted from the frequencies below the local proton plasma frequency ($<$1~kHz) to frequencies above 1~kHz in the spacecraft frame. Their occurrence rate and wave amplitude peaks around the Solar Orbiter's perihelion at distances of 0.5~AU and decreases with increasing distances.

\section{Data}

The Solar Orbiter spacecraft \citep{muller20} was successfully launched on Feb 10, 2020. The first mission of the ESA Cosmic Vision program aims to explore the Sun and heliosphere from close up and out of the ecliptic plane. The spacecraft carries six remote-sensing instruments to observe the Sun and the solar corona and four in-situ instruments to measure the solar wind, energetic particles, and electromagnetic fields. One of the four in-situ experiments is the RPW instrument \citep{maksimovic20}. It is designed to measure magnetic and electric fields, plasma wave spectra and polarization properties, the spacecraft floating potential, and solar radio emissions in the interplanetary medium. Three identical RPW electric antennas, with a length of about 6.5 meters, are mounted on the tip of a 1 meter rigid deployable boom. Antenna V1 is parallel with the spacecraft Z-axis, and antennas V2 and V3 are placed in the Y-Z spacecraft frame (SRF) plane with angles about 125 degrees on both sides from the Z-axis \citep[see Fig. 7 in][]{maksimovic20}. In the SRF frame, the X-axis points
toward the Sun, the Y-axis is the transverse axis, pointing towards the Service Module panel of the spacecraft, that is, closely aligned to the tangential direction, and the Z-axis completes the right-hand frame pointing close to the normal direction.

The Time Domain Sampler (TDS) subsystem of the RPW measures the electromagnetic field in the frequency range from 200~Hz to 200~kHz. The instrument digitizes analog signals from the RPW antennas and the high-frequency winding of the SCM search coil. The RPW/TDS waveform snapshots of the electric component of the electromagnetic field are typically collected from three TDS channels at 262 or 524~kHz sampling rates. In each TDS channel, various configurations of monopole and dipole antenna measurements can be digitized. There are two types of waveform snapshots recorded by the RPW/TDS receiver. Regular survey waveform snapshots (TDS-SURV-RSWF) are taken periodically with a five minute cadence and a typical length of 16 or 32 milliseconds. These snapshots often capture only noise and, thus, they are not used for the present study. Second, triggered survey snapshots (TDS-SURV-TSWF) are on-board selected waveforms based on their intensities and spectral properties. An on-board algorithm analyzes one snapshot with a typical length of 62~ms every second and can efficiently identify coherent waves, such as ion-acoustic or Langmuir waves, which is periodically transmitted in the form of wave and dust counts and average values of relevant parameters. The on-board algorithm also calculates also calculate statistics (TDS-SURV-STAT) of the observed snapshots, and these are transmitted in the form of average values. Parameters included in statistics include peak and RMS amplitude of snapshots and identified waves, wave frequency, a number of identified waves and dust spikes, and the amplitude and width of identified dust spikes. The in-flight performance and a more detailed description of the baseline algorithm is given in \cite{soucek21}. Remaining subsystems of the RPW instrument cover the frequency range of the electric field from DC to 16~MHz and magnetic field from DC up to 200~kHz. Moreover, the BIAS unit samples the spacecraft floating potential and TNR-HFR analyzes thermal noise to provide electron density estimates.

For this study, three electric TDS channels were available and employed. Most of the observations (94\%) were in the dipole mode, with two dipoles (V1-V3 and V2-V1) and one monopole (V2) antenna. Since the RPW antennas are oriented in the Y-Z~SRF plane, namely, the plane perpendicular to the Sun's direction, only two components ($\mathrm{E_Y}$ and $\mathrm{E_Z}$) of the real 3D E-field were used. Electric fields were calculated from two dipole antennas with the effective lengths of $\mathrm{L_{13}=L_{21}=7.53~m}$ \citep[Fig. 9 in ][]{maksimovic20}, as obtained from the ground testing and simulations. The first observations and performance of the RPW instrument with a summary of the most important instrumental noise are reviewed in \cite{maksimovic21}. These interferences do not affect studied waves. An active current bias is applied to each RPW antenna to bring their potential close to the local plasma potential. The antenna configuration and procedure of bias currents calculation is presented in \cite{khotyaintsev21}. They also demonstrated that the measured probe-to-probe potential can be used for electron density measurements with high accuracy. Steinvall et al. (2021) compared the measured electric field provided by RPW with the induced field from $E=-v \times B$ \citep{mozer20a}, where $v$ is the independent solar wind speed estimates by SWA/PAS and $\vec{B}$ is the magnetic field sampled by RPW/SCM. These observations show a good agreement with the directly measured and calculated electric field. Antenna effective lengths used in this study are in the range of experimentally estimated lengths of \cite{steinvall21}.

The ambient magnetic field in the solar wind is measured by the MAG instrument \citep{horbury20}. In the survey mode, MAG operates at the 8~Hz cadence and collects all three magnetic components transformed into the SRF frame. The 1 second averaged magnetic field projected into the Y-Z SRF plane was used to investigate the wave polarization. The solar wind bulk velocity vector, proton density, and temperature are calculated from the Proton and Alpha Sensor (PAS), a part of the Solar Wind Analyzer (SWA) \citep{owen20}. These estimates are available with a typical cadence of 4~Hz from mid-July to mid-October. Both MAG and SWA/PAS data are available through the Solar Orbiter Science Archive \footnote{http://soar.esac.esa.int}.
Electron moments can be derived from the SWA Electron Analyzer System (EAS). Both EAS sensors register the number of electrons per energy and solid angle, from which we can derive the solar wind plasma electron velocity distribution function (eVDF). In this study, we excluded measurements obtained in energies $<$10~eV to avoid the photo-electrons produced on the spacecraft body and accelerated into the instrument by the spacecraft potential. Additionally, we excluded supra-thermal electrons, registered in energies $>$68~eV. Using the eVDF estimates from both sensors, the full 3D velocity distribution function of solar wind electrons was constructed in the spacecraft frame \citep{nicolaou21}. For the purpose of the wave propagation study presented in Section~4, we estimated the electron temperatures for the observations on Oct 14, 2020 (see Fig. \ref{fig:tds_stat_day}).
\section{Wave observations}
\begin{figure}
\centering
 \includegraphics[width=\hsize]{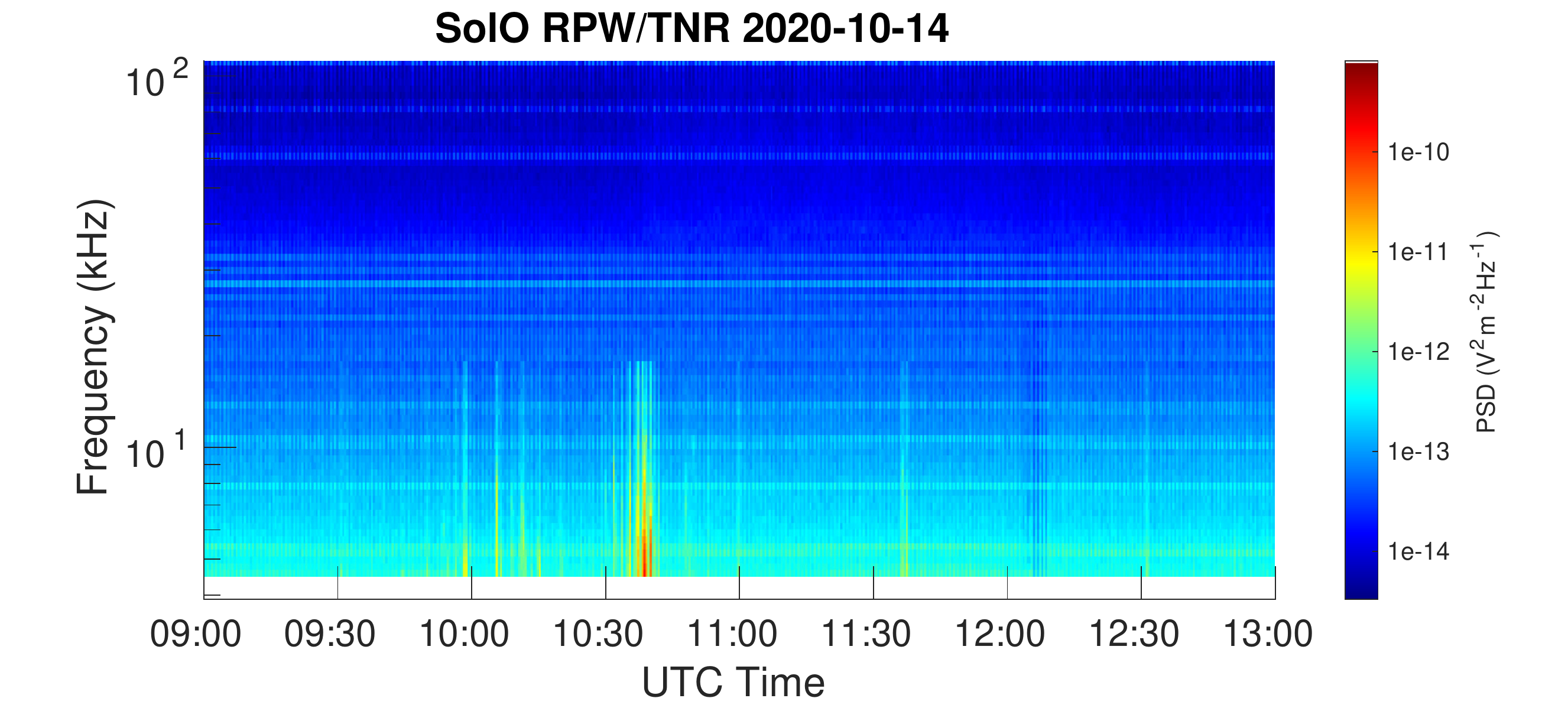}
      \caption{Time-frequency spectrogram with two bursts of intense waves below 20~kHz captured by the RPW/TNR receiver on 14 Oct, 2020.}
         \label{fig:tnr_spec}
  \end{figure}

\begin{figure*}
   \centering
      \includegraphics[width=\textwidth]{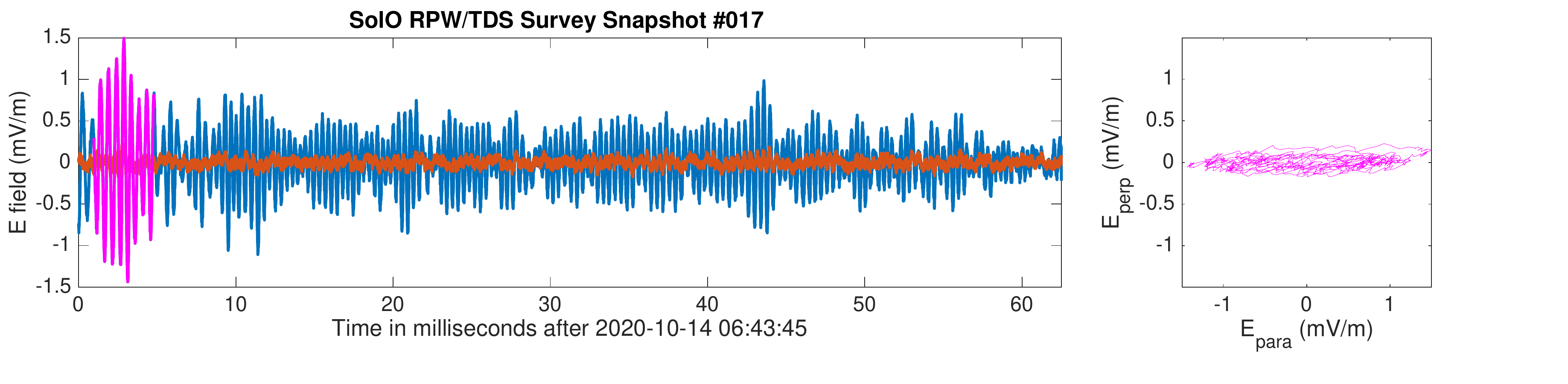}\\
        \includegraphics[width=\textwidth]{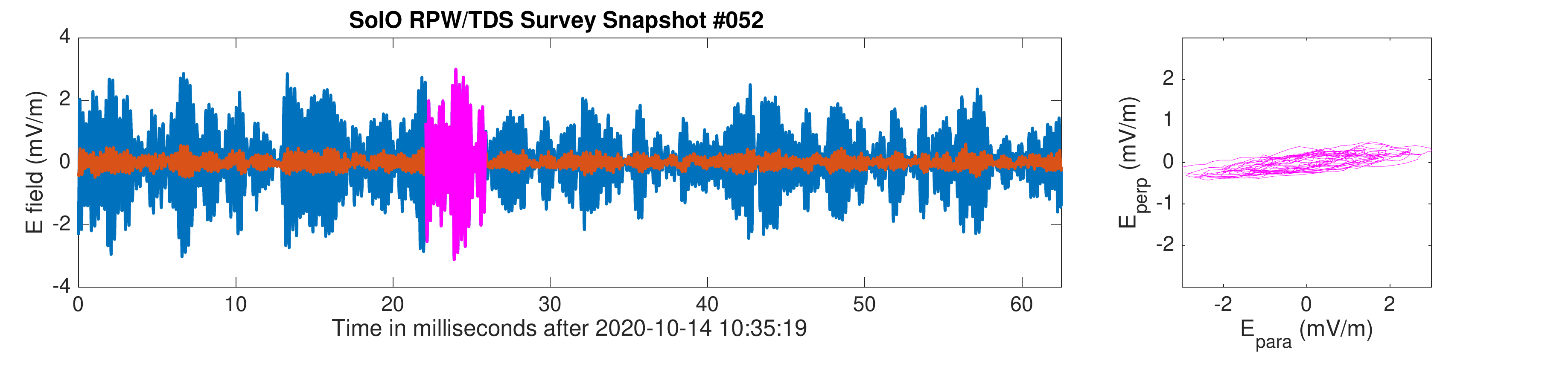}
      \caption{Waveform snapshot with intense waves recorded by the RPW/TDS receiver. The electric field is transformed into the parallel ($\mathrm{E_{para}}$ in blue) and perpendicular ($\mathrm{E_{perp}}$ in orange) directions with respect to the projected ambient magnetic field. In the right column, hodograms of the electric field for the most intense parts of snapshots (in magenta) are shown.}
         \label{fig:tds_snapshots}
  \end{figure*}

After the mission's launch in February 2020 and early commissioning phase in March, the RPW/TDS instrument started its scientific operation in April. Electric field observations often exhibit very intense bursty emissions at frequencies below 20~kHz. Figure~\ref{fig:tnr_spec} presents an example of such emissions captured by the Thermal Noise Receiver (RPW/TNR) between 5 and 100~kHz on Oct 14, 2020. On the time-frequency spectrogram, we can see two bursts of intense waves at frequencies below 20~kHz at about 10:00 and 10:30--10:45~UT. These emissions are at frequencies that are well below the plasma frequency, which was estimated from the RPW/TNR observation to be around 44~kHz, but remain above the electron cyclotron frequency of 150~Hz.
Examples of waveform snapshots with intense waves ($>$1mV/m) from the same time interval are shown in Figure \ref{fig:tds_snapshots}. They show two electric field components parallel (blue) and perpendicular (orange) with respect to the magnetic field direction projected on the Y-Z SRF plane. On the right side, hodograms for the most intense part of the waveform (in magenta) are plotted. Both highlighted wave packets are linearly polarized with a polarization axis parallel to the magnetic field.
  
\begin{figure*}
   \centering
   \includegraphics[width=\textwidth]{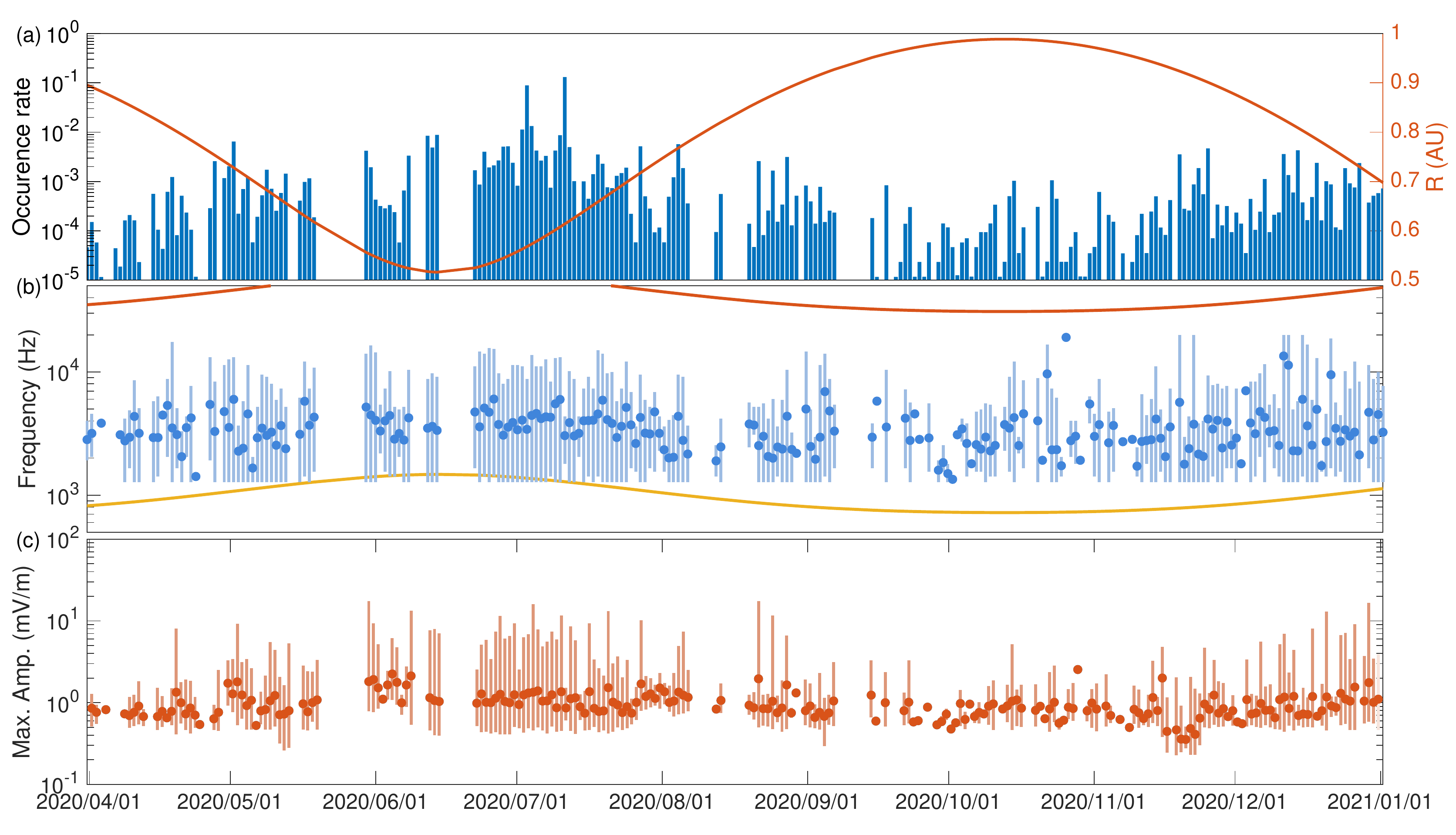}
      \caption{Statistics for the first year of wave detection by the RPW/TDS. Panel~(a) shows an occurrence rate of intense waves below 20~kHz (blue bars) with overplotted distances from the Sun (orange line). A distribution of observed wave frequencies with their variation is shown in the panel~(b). Orange and yellow lines present modelled electron and proton plasma frequency, respectively. Averaged maxima of wave amplitudes with their variations are in the bottom panel~(c).}
         \label{fig:tds_stat_year}
  \end{figure*}

The RPW/TDS statistical data are continuously recorded with a cadence of 16 seconds and allows for an overview of the wave activity at frequencies between 200~Hz and 200~kHz. These first-year statistics are shown in Figure \ref{fig:tds_stat_year}. We excluded days when the instrument detected strong interference above 100~kHz or when the BIAS sweep was in operation. The Venus flyby on Dec~27, 2020 is analyzed in more detail by \cite{hadid21} and is also excluded from the data set here. There were also a couple of days when the receiver was switched off. The occurrence rate has a maximum at distances of about 0.5~AU and reaching values of more than 10\% of daily measured TDS snapshots. The top panel shows an occurrence rate of wave snapshots with detected waves below 20~kHz and normalized to the total valid snapshots recorded per day.  With increasing distances, the occurrence decreases to only a few waves detected per day in October at 0.9~AU. The distribution of wave frequencies calculated as a daily mean (blue circles) with minimal and maximal frequency for a particular day covers the range between 1 and 20~kHz in the middle panel. The wave frequencies mostly occur between electron and proton plasma frequency estimated from the fit of $1/R^2$ model on the SWA/PAS data. The bottom panel presents a daily mean wave amplitude (orange circles) with minimal and maximal amplitudes observed for each day. Wave amplitudes are highly variable during the whole year and reach levels above 10~mV/m. The measured amplitudes show higher mean values and a larger range of obtained values at closer distances from the Sun.

\begin{figure}
   \centering
   \includegraphics[width=.5\textwidth]{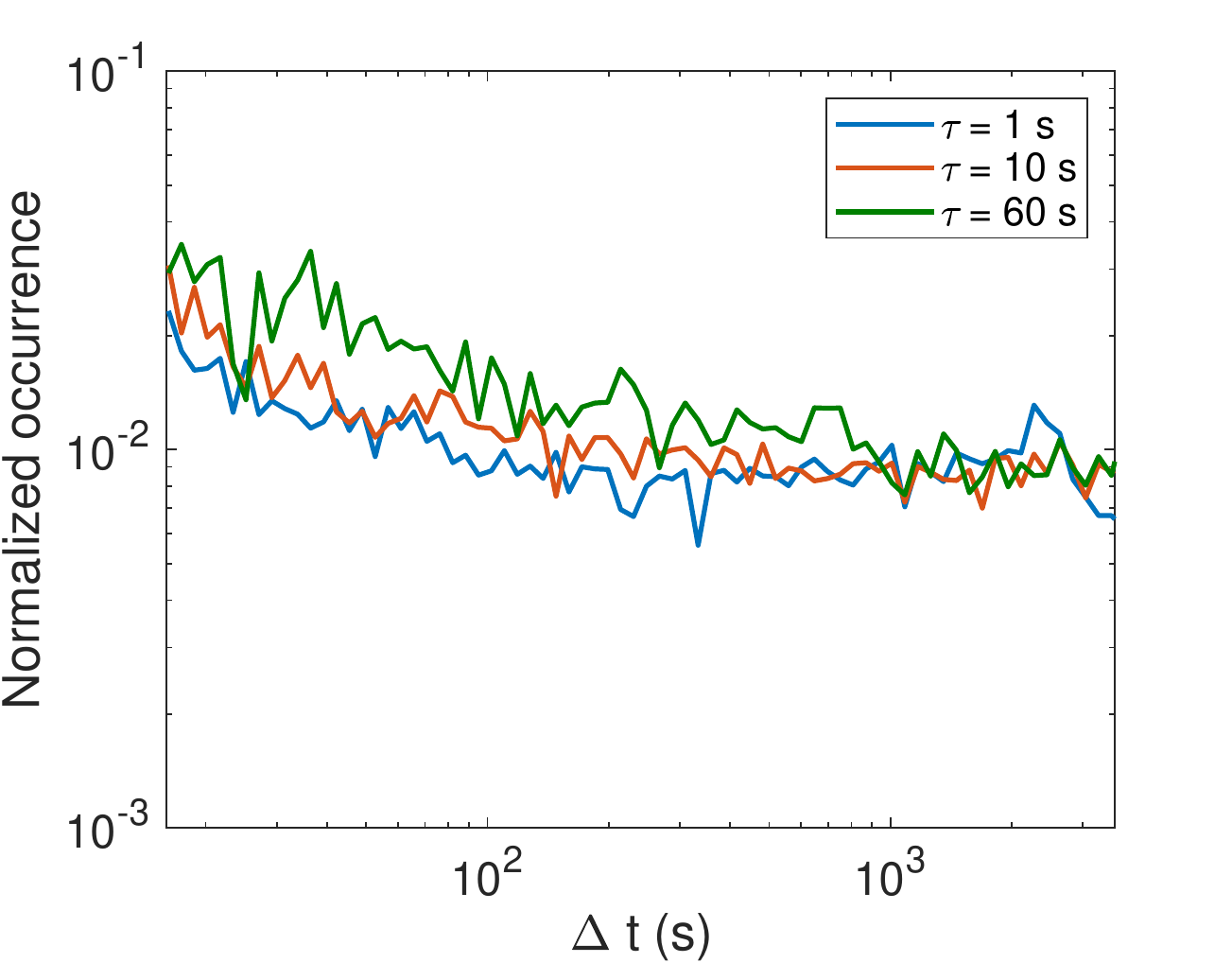}
      \caption{Occurrence rate of time differences, $\Delta t$, for ion-acoustic wave observation and the nearest strong magnetic structure identified by PVI$>$5 for time separations of $\tau $=1~s (blue), $\tau $=10~s (red), and $\tau $=60~s (green).}
         \label{fig:tds_stat_pvi}
  \end{figure}

For a study of magnetic field discontinuities with a connection to observed waves, the Partial Variance of Increments (PVI) method \citep{greco18} was used. The method calculates the vector magnetic field increment along a satellite trajectory as 
\begin{equation}
 \Delta \vec{B}(t,t+\tau)=\vec{B}(t+\tau)-\vec{B}(t),
\end{equation}
where $\vec{B}$ is the magnetic field vector and $\tau$ is a separation in time. Then PVI is defined as the normalized quantity:
\begin{equation}
        \label{eq:pvi}
        \mathrm{PVI} = \frac{|\Delta \vec{B}(t,\tau)|}{\sqrt{\langle | \Delta \vec{B}(t, \tau)|^2\rangle}},
        \end{equation} 
where $\langle \bullet \rangle$ indicates indicates an average over a predefined time interval $\mathrm{T_a >> \tau}$. The PVI method is sensitive to any changes in a direction or magnitude and any form of a sharp gradient in the vector magnetic field and points to phenomena such as switchbacks or shock crossings. The PVI for all available magnetic field data from June to December 2020 was calculated using $\mathrm{\tau}$=1~s, 10~s, and 60~s separations in time and $\mathrm{T_a}$ = one month of data for an average. According to \citep{graham21}, these time separations identify ion-scale structures (1~s) and larger MHD structures (10~s and 60~s) in the solar wind. As a threshold for a larger magnetic field variance, PVI$>$5 was selected. For each separation time, $\mathrm{\tau}$, this threshold exceeds no more than 0.5\% data points. For all TDS statistical observations with almost continuous coverage at 16~s cadence, the closest magnetic structure with PVI$>$5 was found. The distribution of normalized occurrence of time differences, $\mathrm{\Delta t}$, between wave observation and larger magnetic structure for three separation times is in Figure \ref{fig:tds_stat_pvi}. The normalized occurrence is calculated as a ratio of events with low-frequency waves and all available TDS observations in the time difference bin for each time difference. The counts at the smallest time bin (=16~s) correspond to the TDS data observed at or within larger magnetic structures. About 3\% of TDS observations with ion-acoustic waves are observed at or within a magnetic structure for all separation times. Then the occurrence rate is decreasing with increasing time difference to $\Delta t\sim$100~s and remains almost uniform at about 1\% for the remaining time differences.

We used the triggered waveform snapshots from June 2020 to January 2021 for a more detailed wave polarization analysis. We included only those snapshots labeled by the on-board algorithm as waves or unclassified. Snapshots labeled as a dust spike were excluded. Each waveform is transformed from the antenna reference frame into the Y-Z SRF frame. Then auto- and cross-correlations in the form of spectral matrices are calculated \citep{santolik03}. Snapshots with an intense spectral peak ($>$10dB) and coherence greater than 0.8 are taken into account here. For the orientation of the polarization axis, the averaged magnetic field projected into the Y-Z SRF plane was used. Using the criteria above and all TDS measurement in the triggered mode, more than 14000 waveform snapshots with intense wave activity at the frequency range of 200~Hz and 20~kHz were detected. Misidentified waveform snapshots containing dust impacts or solitary structures, typically with higher amplitudes ($>$20~mV/m) and measurements with no simultaneous magnetic field observations were excluded. Since only the projected electric field is observed and assuming electrostatic waves, namely, oscillation along the magnetic field, measured wave amplitudes were multiplied by a factor of $\mathrm{1/sin(\theta_{Bx})}$, where $\mathrm{\theta_{Bx}}$ is an angle between the X SRF and magnetic field direction.
   \begin{figure*}
   \centering
   \includegraphics[width=\textwidth]{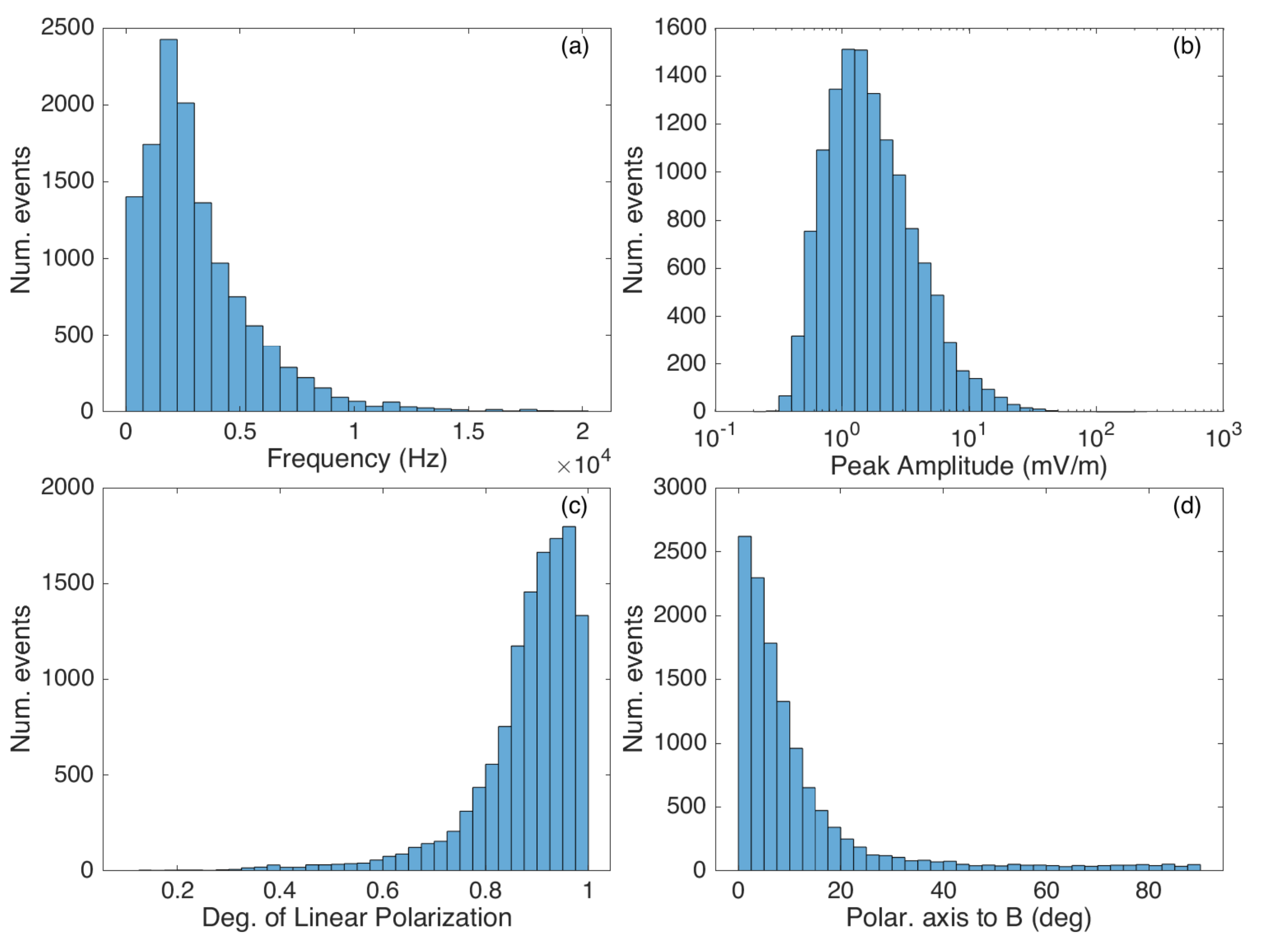}
      \caption{Spectral and polarization analysis of ion-acoustic waves obtained from the RPW/TDS triggered snapshots: (a) Distribution of the peak frequency; (b) Distribution of peak amplitudes; (c) Degree of linear polarization; (d) Angle between wave polarization axis and the projected magnetic field line.}
         \label{fig:tds_tswf_stat}
  \end{figure*}

Figure \ref{fig:tds_tswf_stat} shows the results of a statistic derived from more than 12000 triggered waveform snapshots. The distribution of peak frequency (Fig.\ref{fig:tds_tswf_stat}a) has a maximum of around 2~kHz with four~events below 500~Hz and 61 detected events above 15~kHz. The mean and median frequencies are 3.1 and 2.3~kHz, respectively. Peak amplitudes (in \ref{fig:tds_tswf_stat}b) range from 0.2~mV/m to more than 10~mV/m. The algorithm threshold ($>$10dB above the background) sets the peak amplitude lower limit to 0.2~mV/m. There are 363~events with an amplitude higher than 10~mV/m. The degree of linear polarization calculated from spectral matrices \citep[Eq. 15 in][]{taubenschuss19} shows that more than 80\% of the snapshots have values higher than 0.8. Using the projected magnetic field into the SRF and orientation of the semi-major axis of polarization ellipse \citep[Eq.~11 in][]{taubenschuss19}, a relative angle can be calculated. The distribution of this angle (in panel~d) shows that more than 80\% of the events have an angle that is smaller than 20~degrees from the ambient magnetic field line.

\section{Ion acoustic waves}
RPW/TDS covers frequencies (200~Hz--100~kHz) in the range from the electron cyclotron ($\mathrm{f_{ce}}$) to above plasma frequency ($\mathrm{f_{ce}}$) in heliospheric distances of 0.5--1.0~AU. In this frequency range, only two electrostatic wave modes with wave vector along the magnetic field line exist. Close to the local electron plasma frequency, plasma oscillations or Langmuir waves can occur. For long wavelengths, the electron plasma oscillations are almost purely electrostatic at the electron plasma frequency. As their wavelengths decrease, approaching the Debye length, the frequency of the waves rise above $\mathrm{f_{pe}}$. However, in this region, the oscillations begin to be strongly damped by Landau damping. Electron plasma oscillations are driven via beam instability and often accompanied by the solar wind with Type~III solar radio bursts \citep{gurnett76}. These electron plasma oscillations are rarely observed below 20~kHz.

The ion-acoustic mode can occur at frequencies below electron plasma frequency and they are electrostatic waves generated by the resonant interaction with ion beams or by the current-driven instability. The waves are dispersive, with their phase velocity depending on both electron and proton temperatures. The ion-acoustic mode is strongly damped by Landau damping unless the temperature ratio is $\mathrm{T_e/T_i>1}$, where  $\mathrm{T_e}$ and $\mathrm{T_i}$ are the electron and proton proton temperature, and for wavelengths shorter than the Debye length. A dispersion relation for non-zero electron and proton temperatures can be expressed as follows \citep[Eq. 4.141 in][]{swanson03}:
\begin{equation}
 \label{eqn:dispersion}
        \omega^2_{pl} = \frac{k^2C_s^2}{1+k^2\lambda_{De}^2}\left[1+\frac{3T_i}{T_e}(1+k^2\lambda_{De}^2)\right]
,\end{equation}
where $\mathrm{\omega_{pl}}$ is a wave frequency in the rest frame, $\mathrm{C_s = \sqrt{k_bT_e/m_i}}$ is the ion sound speed , $\mathrm{\lambda_{De}}$ is the electron Debye length, and $\mathrm{k}$ is the wave vector. The Doppler-shifted frequency observed in the spacecraft frame can be calculated as follows:
\begin{equation}
        \label{eqn:dispersion_doppler}
        \omega_{sc} = \omega_{pl} + \vec{k}\cdot\vec{V_{SW}}
        ,\end{equation}
where $\mathrm{\omega_{pl}}$ is the wave frequency in the plasma frame, $\mathrm{\vec{k}}$ is the wave vector, and $\mathrm{\vec{V_{sw}}}$ is the solar wind velocity vector.

\begin{figure}
 \centering
 \includegraphics[width=.5\textwidth]{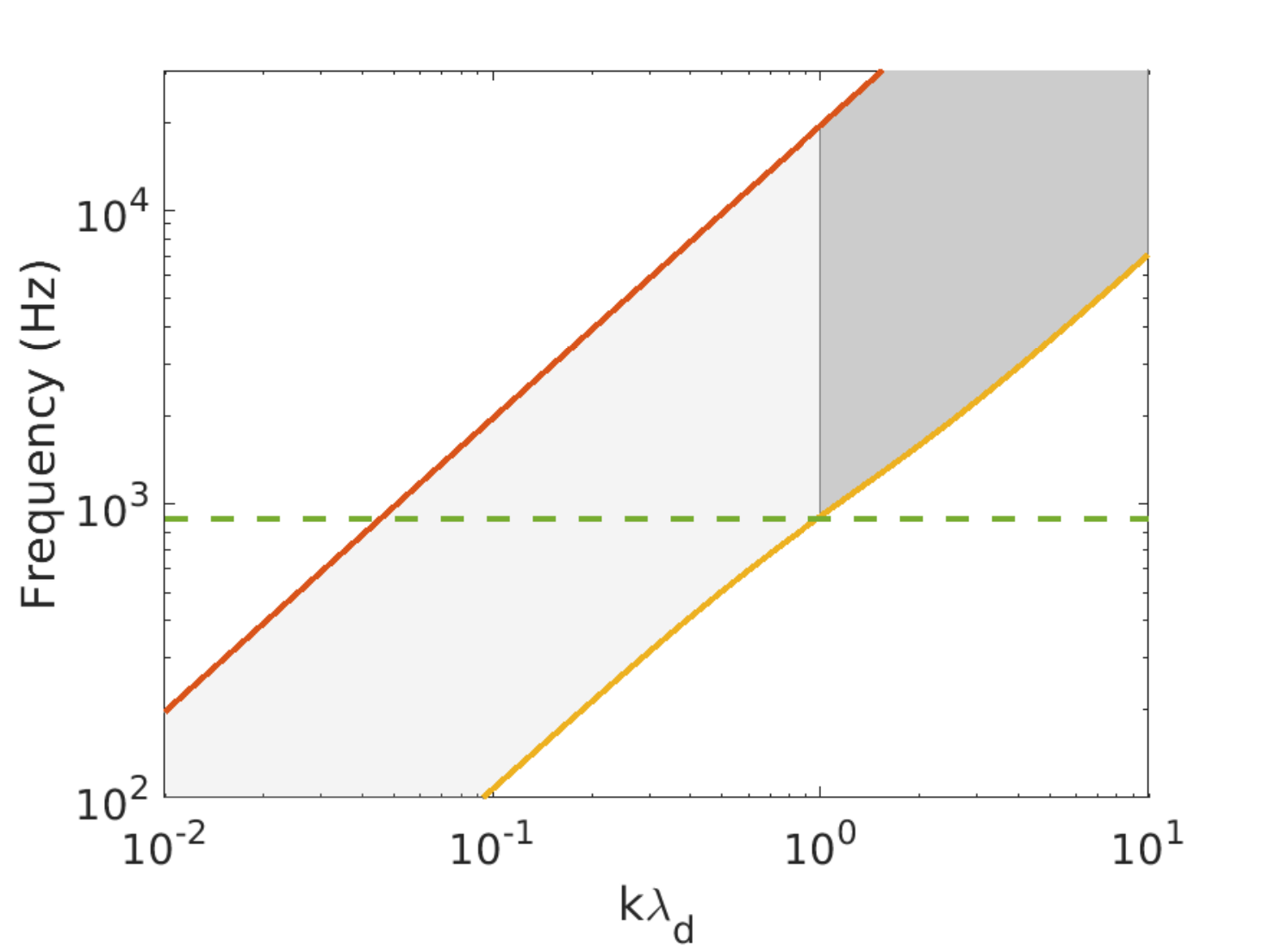}
 \caption{Dispersion relation (yellow line) of ion-acoustic mode for $\mathrm{T_e}$=20~eV, $\mathrm{T_e/T_i}$=4, and $\mathrm{n=15\,cm^{-3}}$ in the plasma rest frame. The red line shows the upper limit of Doppler-shifted frequencies of ion-acoustic waves in the spacecraft frame for the waves propagating parallel or anti-parallel with the solar wind direction with $\mathrm{|V_{sw}|=1000\,km/s}$. The grey area presents typical frequencies of ion-acoustic waves observed in the spacecraft frame. Darker grey delimits a region where ion-acoustic waves start to be strongly damped ($\mathrm{k\lambda_D>1}$). The local proton plasma frequency is shown by the green dashed line.}
 \label{fig:iaw_dispersion}
\end{figure}

A dispersion relation for the ion-acoustic mode following Eq.~\ref{eqn:dispersion}, with $\mathrm{T_e}$=20~eV, $\mathrm{T_i}$=5~eV, and n=$\mathrm{15\,cm^{-3}}$ is plotted in Figure~\ref{fig:iaw_dispersion} as a yellow line. The upper limit of the Doppler-shifted frequencies (Eq. \ref{eqn:dispersion_doppler}) is calculated for a wave vector parallel to the extremely fast solar wind flow with a speed of 1000~km/s \citep{li16}. The grey region delimits all possible frequencies for ion-acoustic mode. The region of strong damping ($\mathrm{k\lambda_{De}>1}$) is indicated by darker grey. The figure shows that a wide range of wave-vectors can be shifted to higher frequencies (1--10~kHz) and above the proton plasma frequency due to the Doppler shift. The mean proton plasma frequency estimated from all SWA/PAS proton density observations is 960~Hz. 

\begin{figure*}
   \centering
   \includegraphics[width=.9\textwidth]{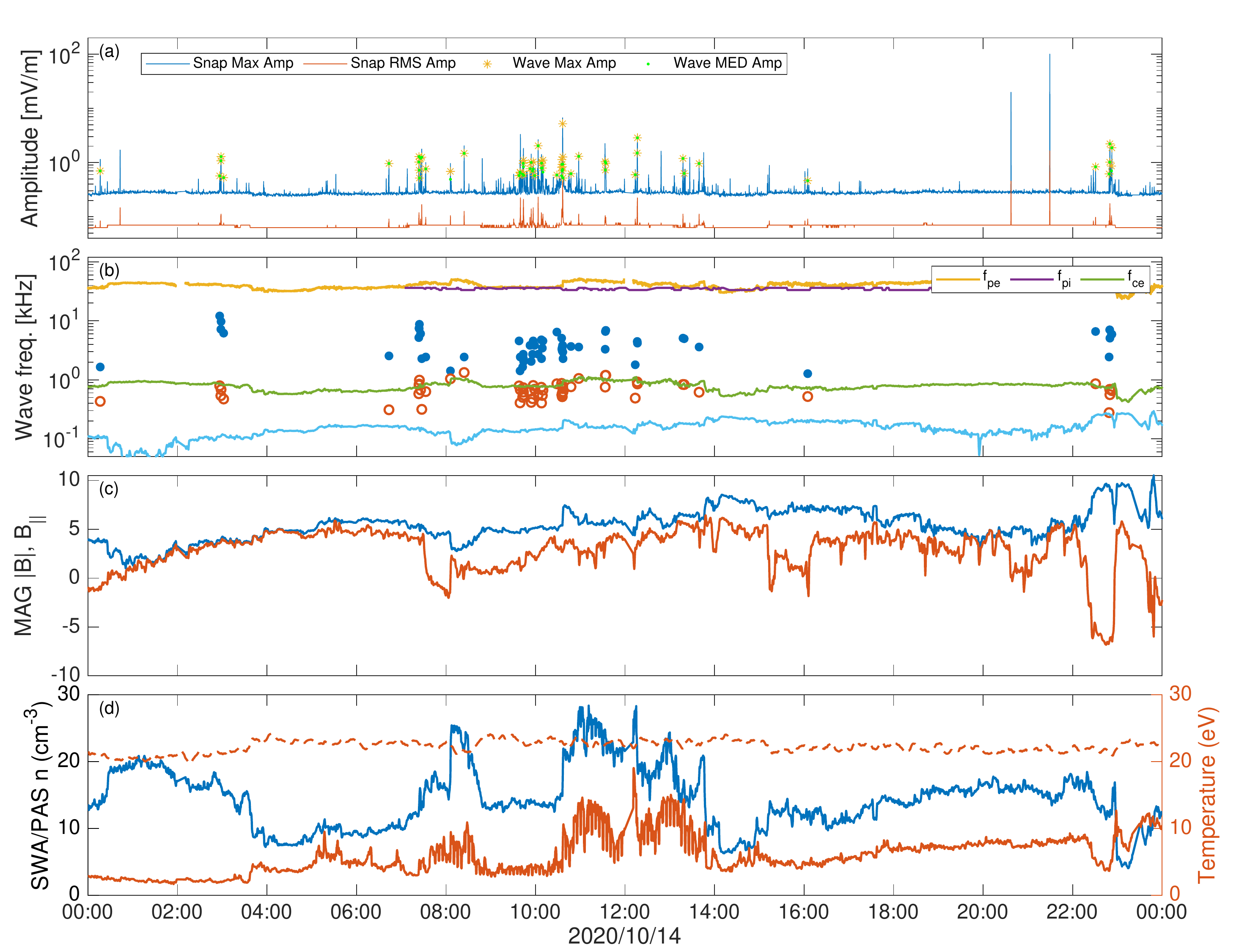}
      \caption{One day of fields and particle observations on Oct 14, 2020: (a) Statistics of RPW/TDS snapshots and wave amplitudes are plotted on the top panel; (b) Distribution of frequencies for detected waves in spacecraft (blue circles) and plasma rest (orange circles) frames. Electron (orange line) and proton (yellow line) plasma frequency estimated from the measurements of RPW/BIAS and SWA/PAS instruments, respectively; (c) Magnitude (blue) and parallel component with respect to the solar wind direction (orange) of the ambient magnetic field from the MAG instrument; (d) Proton density (blue line), proton temperature (orange solid line), and electron temperature (orange dashed line) estimated from the SWA/PAS and SWA/EAS data.}
         \label{fig:tds_stat_day}
  \end{figure*}

A one day statistics of electric and magnetic fields measured on Oct 14 2020, along with the particle data, is in Figure~\ref{fig:tds_stat_day}. The statistics of peak and root mean square amplitudes from TDS snapshots (blue and red lines) and identified waves (yellow and green crosses) show higher wave activity between 08:00 and 14:00 in panel~(a). While snapshot statistics is continuous with a cadence of 16 seconds, waves need to be identified by the on-board algorithm to be stored. There are also two dust spikes between 20:00 and 22:00 with amplitudes $>$10~mV/m. Wave frequencies cover the range of 1--10~kHz (blue circles) and fit between the plasma frequencies derived from the RPW/BIAS observations (yellow line) and proton plasma frequencies estimated from the SWA/PAS data (violet line) in the panel~(b). Using the one-minute average of solar wind plasma parameters, magnetic field, and the dispersion relation from Eq.~\ref{eqn:dispersion}, the wave frequency in the plasma rest frame was estimated (open orange circles). The dispersion relation gives us up to three possible wave vectors depending on their orientation (parallel or anti-parallel) to the solar wind flow and wave frequency in the spacecraft frame. The solution of the dispersion relation with the lowest wave vector was preferred due to the expected lower attenuation. The rest frame wave frequency almost fits between electron cyclotron frequency (green line) and proton plasma frequency (violet line). The magnitude (blue line) and projection on the solar wind direction of the magnetic field observations are shown in the  panel~(c). The bottom panel~(d) shows a one-minute average of proton density (blue line) and temperature (orange line) calculated from the SWA/PAS observations. The orange dashed line represents the electron temperature calculated from the SWA/EAS measurements. Wave activity is associated with significant changes in the magnetic field line configuration. These abrupt changes are also evident in proton observations. 

\section{Discussion}
The electric field measurements of the RPW/TDS receiver often capture very intense broadband fluctuations at frequencies below 20~kHz in the spacecraft frame. Using the RPW/TDS statistical data based on the continuously captured processed properties of the waveform snapshots, we studied the plasma waves at frequencies below 20~kHz that cover an interval of heliocentric distances between 0.5 AU and 1~AU. The wave occurrence rate peaks close to the first perihelion at distances of about 0.5~AU and reaching more than 10\% of all downlinked triggered snapshots. With increasing distances, the occurrence rate decreases to only a few waves detected per day. The distribution of peak frequencies covers the range from 1 to 10~kHz and with typical amplitudes of 1--10~mV/m, with higher values close to the perihelion. 
\begin{figure}
   \centering
   \includegraphics[width=\hsize]{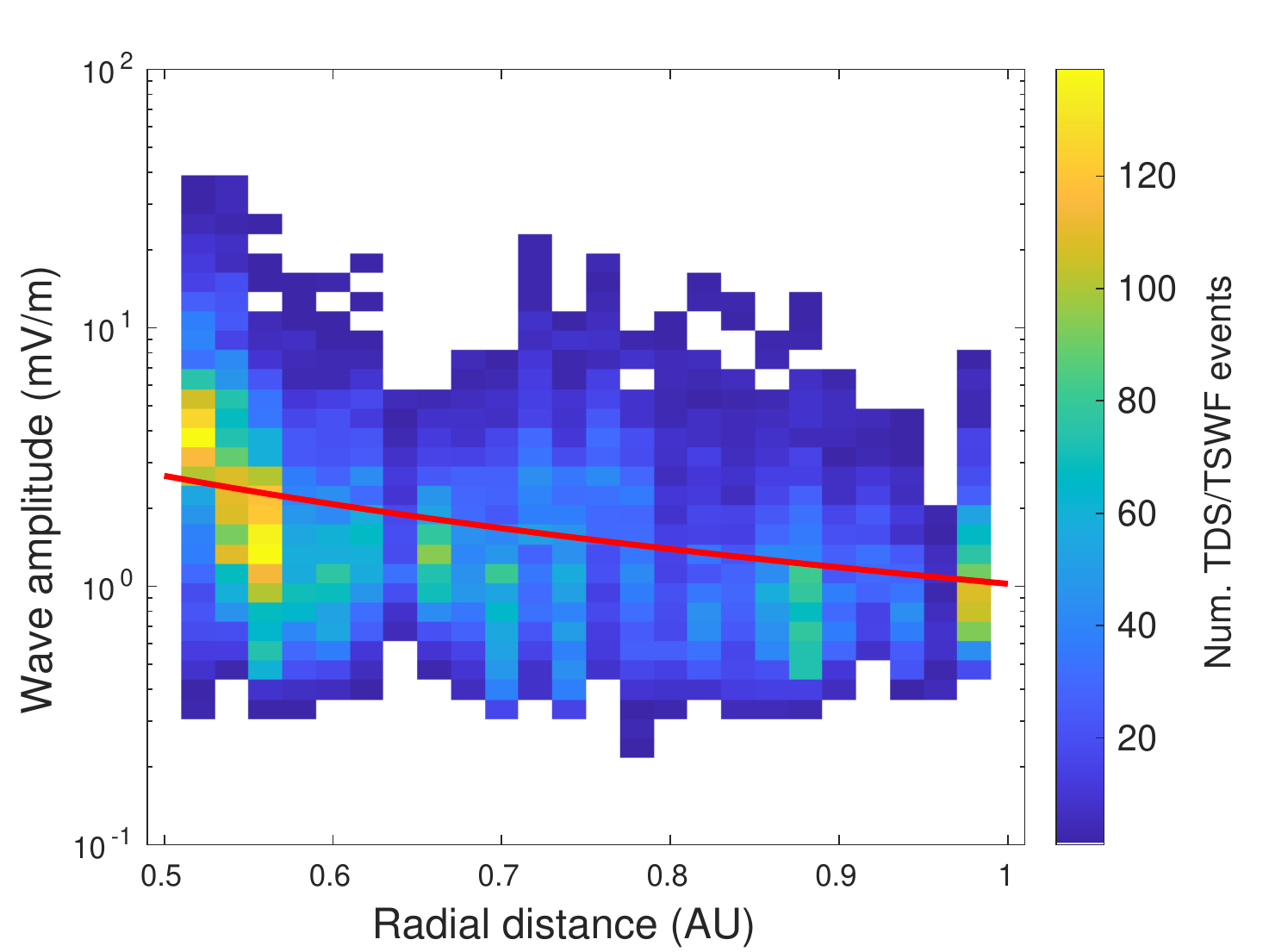}
      \caption{Wave amplitudes of ion-acoustic waves captured by the triggered snapshots and corrected for the projection to Y-Z SRF plane as a function of heliocentric distances between 0.5 and 1.0 AU. The red line shows the least square fit of wave amplitudes as $E \propto R^{\alpha}$ with $\alpha=-1.39\pm0.33$.}
         \label{fig:tds_tswf_emax}
  \end{figure}

A more detailed analysis of the triggered waveform snapshots shows the mean wave frequency about 3~kHz and wave power about $\mathrm{5\cdot10^{-1}~mV^2/m^2}$. The distribution of wave amplitudes varies from 0.2 to 20~mV/m with a mean value of 2.5~mV/m. As shown in Figure~\ref{fig:tds_tswf_emax}, wave amplitudes decrease with increasing distance from the Sun. The least-square fit shows the wave amplitude dependence as $\mathrm{R^{-1.39}}$. The relative phase distribution between two components of the E-field shows a mostly linear wave polarization, while electric field fluctuations are closely aligned with the directions of the ambient magnetic field with the mean value of 12~degrees. The analyzed waves are interpreted as a strongly Doppler-shifted electrostatic ion-acoustic mode.  

Earlier spectral measurements on the Parker Solar Probe \citep{mozer20b} showed large-amplitude wave bursts identified as Doppler-shifted ion-acoustic waves at heliocentric distances of 55~$\mathrm{R_S}$ (0.26~AU). More than one-third of captured wave bursts were measured inside the magnetic field switchbacks \citep{dudokdewit20,laker20}. 

\cite{mozer20b} suggest that ion acoustic waves are generated by the ion-beam instability and are a general feature of switchbacks that occur frequently near the Sun. Our observations in Figure \ref{fig:tds_stat_day} shows wave observation during the time period when the magnetic field direction is highly disturbed. The statistical analysis of magnetic field discontinuities with a connection to observed waves showed that about 3\% of the events are within magnetic discontinuity for all three separations in time (Fig. \ref{fig:tds_stat_pvi}). These results show that there is a small increase of occurrence rate of ion-acoustic waves  associated with magnetic discontinuities. 

The ion-acoustic mode may also grow from a current-driven or heat flux instability \citep[e.g.,][]{forslund70,lemons79}. The threshold for this instability is rather high for the ratio $\mathrm{T_e/T_i}\sim{1-2}$ that is typical of the solar wind at 1~AU \citep[][ and references therein]{wilson18}. The drift velocity of the electrons with respect to ion velocity should be about the electron thermal velocity in order for the wave growth to occur. Our first estimates of electron and proton temperature, observed on Oct 14, 2020, show a higher ratio $\mathrm{T_e/T_i\sim4}$, corresponding to a minimum drift velocity of $\mathrm{\sim0.3C_s}$, where $\mathrm{C_s}$ is the electron sound velocity \citep[Fig. 9.32 in ][]{gurnett17}. The higher temperature ratio ($\mathrm{T_e/T_i\sim2-6}$) was also presented in observations from ACE \citep{skoug00} or Helios \citep{march82} spacecraft. This higher temperature ratio is more favorable and decreases the threshold for current-driven instability. The analysis of the associated currents observed by Solar Orbiter during the current sheet crossings by \cite{graham21} shows that the observed currents are unlikely to provide wave growth via the current-driven instability. The calculated threshold currents would still be well above any observed currents. These authors conclude that although the waves are more likely to be found in enhanced current regions, the current-driven instability cannot generate the waves. On the other hand, \cite{wilson20} found evidence of above-threshold drift speed between electron and ions populations for almost 5\% events observed by the Wind spacecraft at interplanetary shocks. They also noted that the presence of gradients in the temperature and density reduces the idealized critical drift by a factor of $\sim2-8$. This effect would increase the relative number of events exceeding the critical drift threshold to almost 10\%. This explanation would be consistent with our observation that the waves in our dataset are observed more frequently in the vicinity of such changes in the plasma parameters. The problem of the unstable electron and ion populations for ion-acoustic waves is, however, rather complex, as discussed, for example, by \cite{dum80}. Even if the stability calculations for the modeled distribution functions give no wave growth, small changes in the shape of the distribution function within the error bars of the observed distributions can often result in instability.

\section{Conclusions}

Electrostatic ion-acoustic waves are the most common wave emissions observed below the local electron plasma frequency by the RPW/TDS receiver during the mission's first year. These waves are measured as intense bursts with amplitudes up to tens of mV/m. The number of observed waves and wave amplitude is increasing with decreasing distances from the Sun. Waves are strongly Doppler-shifted from the frequencies below proton plasma frequency ($<$1 kHz) to frequencies up to 10~kHz in the spacecraft frame. The majority of detected waves have a high degree of linear polarization with the polarization axis aligned with the ambient magnetic field line. Our nearly continuous mapping of ion-acoustic waves shows that they are a common phenomenon at heliocentric distances between 0.5 and 1~AU. Ion-acoustic waves may accompany large-scale solar wind structures, as interplanetary shocks or rotations of the magnetic field (switchbacks) \citep{wilson18,mozer20b}. Our analysis shows that about 3\% of ion-acoustic waves are observed at or within large magnetic structures. This occurrence rate is three times higher than an average value of 1\% for separation times between 16 s and one hour. Thus, ion-acoustic waves are more likely observed within magnetic field discontinuities. Our analysis shows that about 3\% of ion-acoustic waves are observed at or within large magnetic structures. This occurrence rate is three times higher than an average value of 1\% for separation times between 16~s and one day. Thus, ion-acoustic waves are more likely observed within magnetic field discontinuities, but they are not exclusively connected to these processes. The ion-acoustic waves can be generated by the resonant interaction with ion beams or by the current-driven instability. Even if the ion-ion instability appears more plausible \citep{mozer20b,graham21}, conditions that are favorable to the current-driven instability \citep{wilson20} may occur near the plasma and field gradients, in particular, and a contribution on the part of the current-driven instability cannot be ruled out.

\begin{acknowledgements}
Solar Orbiter is a space mission of international collaboration between ESA and NASA, operated by ESA. RPW/TDS development has been developed with the support from the Czech PRODEX program funded by the Czech Ministry of Education, Youth and Sports. The TDS team thanks the scientific and engineering teams or other RPW subsystems as well as the team from the French Space Agency CNES, CNRS, the Paris Observatory, and The Swedish National Space Agency for their support of the TDS development. 
\end{acknowledgements}
%
%

%
	
\end{document}